\documentclass[9pt,conference]{IEEEtran}
\IEEEoverridecommandlockouts

\usepackage{cite}
\usepackage{amsmath,amssymb,amsfonts,graphicx}
\usepackage{algorithmic}
\usepackage{graphicx}
\usepackage{textcomp}
\usepackage{xcolor}
\usepackage{booktabs}
\usepackage{color}
\usepackage{colortbl}
\usepackage{multirow}
\usepackage{multicol}
\usepackage{adjustbox}
\usepackage[flushleft]{threeparttable}
\usepackage[colorlinks,
            linkcolor=red,       
            anchorcolor=blue,  
            citecolor=green,       
            ]{hyperref}
\def\BibTeX{{\rm B\kern-.05em{\sc i\kern-.025em b}\kern-.08em
    T\kern-.1667em\lower.7ex\hbox{E}\kern-.125emX}}
\begin{document}

\title{Generalizable Audio Deepfake Detection via Latent Space Refinement and Augmentation
\thanks{$^{\dagger}$ Corresponding author}
}


\author{
\IEEEauthorblockN{
Wen Huang$^{1,2}$\qquad
Yanmei Gu$^{3}$\qquad
Zhiming Wang$^{3}$\qquad
Huijia Zhu$^{3}$\qquad
Yanmin Qian$^{1\dagger}$}
\IEEEauthorblockA{
$^{1}$\textit{Auditory Cognition and Computational Acoustics Lab,
MoE Key Lab of Artificial Intelligence, AI Institute}\\
\textit{Department of Computer Science and Engineering, 
Shanghai Jiao Tong University}, Shanghai, China\\
$^{2}$\textit{SJTU Paris Elite lnstitute of Technology}, 
$^{3}$\textit{Ant Group}, Shanghai, China}
}

\maketitle

\begin{abstract}
Advances in speech synthesis technologies, like text-to-speech (TTS) and voice conversion (VC), have made detecting deepfake speech increasingly challenging. Spoofing countermeasures often struggle to generalize effectively, particularly when faced with unseen attacks. To address this, we propose a novel strategy that integrates Latent Space Refinement (LSR) and Latent Space Augmentation (LSA) to improve the generalization of deepfake detection systems. LSR introduces multiple learnable prototypes for the spoof class, refining the latent space to better capture the intricate variations within spoofed data. LSA further diversifies spoofed data representations by applying augmentation techniques directly in the latent space, enabling the model to learn a broader range of spoofing patterns. We evaluated our approach on four representative datasets, i.e. ASVspoof 2019 LA, ASVspoof 2021 LA and DF, and In-The-Wild. The results show that LSR and LSA perform well individually, and their integration achieves competitive results, matching or surpassing current state-of-the-art methods.
\end{abstract}


\section{Introduction}
\label{sec:intro}
With advancements in speech synthesis systems such as text-to-speech (TTS) and voice conversion (VC), detecting deepfake speech has become increasingly challenging. Synthesized data can originate from a wide range of synthesis systems, each with its own distinct characteristics, making it difficult for spoofing countermeasures to generalize effectively. This challenge is exacerbated when detectors encounter unseen deepfake attacks, often leading to significant performance degradation~\cite{yamagishi2021asvspoof, muller2022does}.

To enhance generalization in deepfake detectors, one key direction focuses on developing more robust classification models through improved architecture and learning strategies. Recent studies have utilized features extracted from self-supervised speech models such as Wav2vec~\cite{babu2021xlsr}, Whisper~\cite{radford2023robust}, and WavLM~\cite{chen2022wavlm} as front-end inputs for deepfake detection. These models, trained on large-scale and diverse speech data, strengthen the detection process by providing reliable and domain-agnostic features~\cite{tak2022automatic}. Beyond improving feature extraction, researchers have also worked to improve the accuracy of back-end classifiers. Traditional binary classification methods often struggle with generalization, particularly when facing distribution mismatches. To address this, one-class learning approaches have been explored, focusing on creating a compact representation of bonafide speech while effectively pushing away spoofed speech, leading to a well-separated and more generalizable feature space~\cite{zhang2021one, kim2024one}.

Another promising direction is through data augmentation, which enhances the robustness of the model by exposing it to a wider range of data variations during training. Traditional techniques such as speed perturbation, SpecAugment~\cite{park2019specaugment}, and codec augmentation have been shown to improve performance. More recent methods, such as Rawboost~\cite{tak2022rawboost}, use signal processing techniques to boost or distort raw audio, leading to significant improvements. There are also augmentation strategies specifically designed for audio deepfake detection. For instance, CpAug~\cite{zhang2024cpaug} employs a copy-paste strategy to generate diverse training samples, while targeted augmentation methods~\cite{astrid2024targeted} create pseudo-fakes that challenge the decision boundary, thereby increasing the diversity of fake samples. Furthermore, research has shown that using neural vocoders to augment data can further enhance detection performance~\cite{wang2023spoofed, wang2024can}.

\begin{figure*}[ht]
    \centering
    \includegraphics[width=1.0\linewidth]{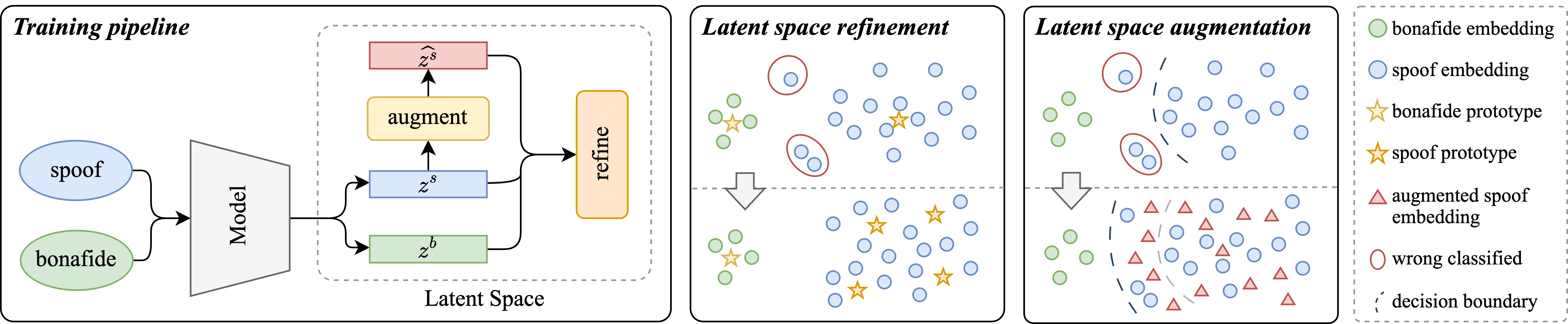}
    \caption{The pipeline of the proposed method, illustrating the process of Latent Space Refinement (LSR) and Latent Space Augmentation (LSA).}
    \label{fig:pipeline}
\end{figure*}

Building on these two key directions, we propose a novel strategy of integrating latent space refinement and augmentation to further boost the generalization ability of deepfake detection, as shown in Fig.~\ref{fig:pipeline}.
First, to address the limitations of binary classification in capturing the diverse nature of spoofed audio, we introduce \textbf{Latent Space Refinement (LSR)}. In binary classification, models typically assign a single prototype to each class, which oversimplifies the complex variability within spoofed audio. While one-class learning tries to address this by compactly representing the bonafide class and treating others as outliers, it often imposes a rigid boundary that fails to capture the diversity in spoofed data.
In contrast, our LSR approach introduces multiple learnable prototypes specifically for the spoof class, refining the latent space to better model the intricate variations within spoofed data. This enhanced representation reduces intra-class variability and allows the model to generalize more effectively across different spoofing attacks.

Second, to further enhance generalization, we apply \textbf{Latent Space Augmentation (LSA)} to diversify spoofed data representations, inspired by successful applications in computer vision~\cite{liu2018data, yan2024transcending}. Unlike traditional data augmentation, which focuses on manipulating input data, LSA directly targets the latent space, allowing it to be independent of specific audio-level operations. By applying techniques such as additive noise, affine transformation, batch mixup, and linear interpolation and extrapolation, LSA generates a wide range of spoofed examples that expand the latent space. This expansion helps the model capture more diverse patterns within spoofed data, thereby improving its ability to generalize across different spoofing attacks and enhancing overall detection performance.

Our experimental results confirm the effectiveness of the proposed latent space refinement and augmentation. We evaluated the approach on four representative datasets: ASVspoof 2019 LA~\cite{wang2020asvspoof}, ASVspoof 2021 LA and DF~\cite{yamagishi2021asvspoof}, and In-The-Wild~\cite{muller2022does}. The findings show that both LSR and LSA individually contribute to performance improvements, with the integrated system achieving competitive results, matching or surpassing the current state-of-the-art across these diverse benchmarks.

\section{Methods}
\label{sec:method}

\subsection{Latent Space Refinement}
To capture the inherent variations within the spoof class, we introduce multiple learnable prototypes that refine the latent distribution. Assume there are $K$ prototypes for each class, denoted as $\{c_1, \dots, c_K\}$. For the bonafide class, $K=1$, while for the spoof class, $K$ is a hyperparameter chosen based on the complexity of the data. To determine the probability of a sample $x$ belonging to a particular class, we compute the maximum cosine similarity between its embedding $z$ and each of the class prototypes
:
\begin{equation}
    \cos\theta = \sum_{i=1}^K\frac{e^{\langle c_i, z\rangle\cdot\gamma}}{\sum_{j=1}^K e^{\langle c_i, z\rangle\cdot\gamma}}\langle c_i, z\rangle
\end{equation}
where $\langle x,y\rangle=\frac{x\cdot y}{\|x\| \|y\|} $ represents the cosine similarity between two vectors, and $\gamma$ is the scaling factor, set to 10. We smooth the maximum operator using a softmax-like operation to prevent sensitivity between multiple prototypes.

To guide the learning of these prototypes, we design a prototype-based classification loss, inspired by the additive angular margin loss~\cite{deng2019arcface}:
\begin{equation}
    \mathcal{L}_{proto}(z)=-\log \frac{e^{s(\cos(\theta_{y+m}))}}{e^{s(\cos(\theta_{y}+m))}+e^{s(\cos\theta_{1-y})}}
\end{equation}
Here, $y\in\{0, 1\}$ is the label of sample $x$, $m$ is an angular margin penalty, and $s$ is a scaling factor. This loss function encourages the model to push the embeddings of genuine samples closer to the bonafide prototype and spoofed samples closer to their corresponding prototypes.

While prototypes are learned during the training process, there’s a risk that they may collapse to a single center. To mitigate this, we introduce an intra-class regularization for the spoof prototypes $\{c^s\}$:
\begin{equation}
    \mathcal{L}_{intra}(\{c^s\})=\frac{2}{K(K-1)}\sum_{i=1}^{K-1} \sum_{j=i+1}^K \langle c^s_i, c^s_j \rangle
\end{equation}
This regularization term calculates the mean similarity between the spoof prototypes, encouraging them to spread out in the latent space, thereby preventing prototype collapse.

To further enhance the distinction between spoof and bonafide prototypes, we introduce an inter-class regularization term. This term calculates the smoothed maximum cosine similarity between the spoof prototypes $\{c^s\}$ and the single bonafide prototype $c^b$:
\begin{equation}
    \mathcal{L}_{inter}(\{c^s\}, c^b)=\delta+\sum_{i=1}^K\frac{e^{\langle c^s_i, c^b\rangle\cdot\gamma}}{\sum_{j=1}^K e^{\langle c^s_i, c^b\rangle\cdot\gamma}}\langle c^s_i, c^b\rangle
\end{equation}
here $\delta$ is a regularization coefficient that prevents the loss from becoming negative. 

Hence, the overall objective function for LSR is defined as follows:
\begin{equation}
    \mathcal{L}_{LSR}=\mathcal{L}_{proto}+\mathcal{L}_{intra}+\mathcal{L}_{inter}
\end{equation}
In addition, the LSR loss can be incorporated alongside a binary classification loss, such as Weighted Cross Entropy (WCE), to refine the latent distribution and reduce intra-class variance.

\subsection{Latent Space Augmentation}
While multi-prototypical refinement enhances the representation of the spoofed class, further generalization can be achieved by augmenting the diversity of the training data. Instead of solely augmenting raw input data, we apply augmentation directly in the latent space, where lower dimensionality allows for more targeted variations. By focusing these augmentations on spoofed latent features, we generate diverse spoofing variations. Notably, these augmentations are not applied to bonafide latent features, preserving their authenticity.

Given $z$ a batch of embeddings, we denote the spoof embeddings in this batch as $z^s$ and the bonafide embeddings as $z^b$. To create diverse variations of spoof embeddings, we design five latent augmentation patterns for $z^s$:

\textbf{Additive Noise (AN).} 
A simple yet efficient idea is to add random perturbation to latent features. Here we apply the additive noise drawn from a Gaussian distribution as follows: 
\begin{equation}
    \hat{z^s}=z^s+\beta\cdot X, X\sim \mathcal{N}(0, \mathbf{I})
\end{equation}
where $\mathcal{N}(0, \mathbf{I})$ is the standard normal distribution, $\mathbf{I}$ is the identity matrix, and $\beta$ is a scaling factor sampled from $\mathcal{N}(0,1)$.

\textbf{Affine Transformation (AT).} 
This common transformation for 1D vectors involves scaling and translating the latent features:
\begin{equation}
    \hat{z^s}=a\cdot z^s + b
\end{equation}
where $a$ is sampled from $\mathcal{U}(0.9,1.1)$ and $b$ is set to 0. 

\textbf{Batch Mixup (BM).} 
Inspired by data mixup strategies~\cite{zhang2017mixup}, we creates new latent features by blending pairs of spoof features in the batch, creating smoother transitions and intermediate variations:
\begin{equation}
    \hat{z_i^s}=\alpha\cdot z_i^s+(1-\alpha)\cdot z_{\pi(i)}^s
\end{equation}
where $i$ indexes the batch, $\pi$ denotes a random permutation of the batch indices and $\alpha$ is a mixup coefficient sampled from $\mathrm{Beta} (0.5,0.5)$. 

The following two techniques rely on the prototypes learned in latent space refinement:

\textbf{Linear Interpolation (LI).} 
To create more challenging examples targeting the decision boundary, we perform linear interpolation on spoof embeddings towards bonafide prototype $c^b$. Since the prototypes in LSR the prototypes are normalized to lie on a unit hypersphere due to the use of cosine similarity, the norm of the vectors is incorporated to adjust for the transition to Euclidean space:
\begin{equation}
    \hat{z^s}=z^s+\lambda_i \cdot(\frac{\|z^s\|}{\|c^b\|}c^b -z^s)
\end{equation}
where $\lambda_i$ is an interpolation coefficient sampled from $\mathcal{U}(0,0.1)$, and the norm term $\|z^s\|/\|c^b\|$ aligns the scales of the vectors.

\textbf{Linear Extrapolation (LE).} 
In addition to interpolation, we also perform extrapolation from the nearest spoof prototype to create new features:
\begin{equation}
    \hat{z^s}=z^s+\lambda_e \cdot(z^s -\frac{\|z^s\|}{\|c^s_n\|}c^s_n)
\end{equation}
where $c^s_n$ corresponds the nearest spoof prototype of $z^s$ and $\lambda_e$ is an extrapolation coefficient sampled from $\mathcal{U}(0,0.1)$. Similarly, we use the norm $\|z^s\|/\|c^s_n\|$ to adjust for the Euclidean representation. This method extends the spoof features further away from the nearest prototype, generating more diverse variations.

Finally, the augmented latent features $\hat{z^s}$ are concatenated with the original features $z$, forming $z'=[z\parallel\hat{z^s}]$. These enhanced features are then used for loss calculation during subsequent training, allowing the model to learn from a more varied set of spoofed data.

\section{Experiments}
\subsection{Experimental Settings}
\label{ssec:set}
\noindent\textbf{Datasets and metrics.}
We train all systems using the ASVspoof 2019 LA training set~\cite{wang2020asvspoof}, which includes approximately 25k utterances and 6 spoofing attacks involving VC or TTS. To evaluate generalization performance, we test on multiple datasets: the ASVspoof 2019 LA evaluation set (\texttt{19LA})~\cite{wang2020asvspoof}, containing 71k utterances with 13 different spoofing attacks; the ASVspoof 2021 LA set (\texttt{21LA})~\cite{yamagishi2021asvspoof}, comprising about 181k utterances with algorithms similar to \texttt{19LA} but also reflecting telephony systems' encoding and transmission effects; the ASVspoof 2021 DF set (\texttt{21DF})~\cite{yamagishi2021asvspoof}, with over 600k utterances and more than 100 spoofing attacks processed with various lossy codecs; and the In-The-Wild dataset (\texttt{ITW})~\cite{muller2022does}, which features approximately 32k utterances collected under real-world, non-controlled conditions, making it a more challenging dataset.
Performance is measured using Equal Error Rate (EER).

\noindent\textbf{Training details.}
We adopt the model architecture from \cite{tak2022automatic}, utilizing Wav2Vec2.0 XLSR~\cite{babu2021xlsr} as the frontend feature extractor and AASIST~\cite{jung2022aasist} as the backend classifier. Input speech is randomly chunked into 4-second segments, with Rawboost~\cite{tak2022rawboost} applied as basic augmentation and codec augmentation as extra augmentation. The learning rate is set to 1e-6 for the backbone model and 1e-3 for the prototypes in LSR. 
For the LSR loss, we set the scaling factor $s=32$, angular margin $m=0.2$, and regularization coefficient $\delta=0.2$. For the WCE loss, the weights for bonafide and spoof classes are set to 0.9 and 0.1, respectively. For LSA, we either fix one type of augmentation during training or randomly select from all augmentation types (denoted as \textit{All}).

\subsection{Overall Performance Comparison}

\begin{table}[htbp]
\centering
\caption{Overall performance comparison in EER(\%) across multiple datasets. All systems are trained on the ASVspoof2019 LA training set. Best results are highlighted in \textbf{bold}, and second-best results are \underline{underlined}.}
\label{tab:compare}
\begin{tabular}{l|cccc}
\toprule
System & \texttt{19LA} & \texttt{21LA} & \texttt{21DF} & \texttt{ITW} \\ \midrule
WavLM+AttM~\cite{pan2024attentive} & 0.65 & 3.50 & 3.19 & - \\
Wav2Vec+LogReg~\cite{oneata2023towards} & 0.50 & - & - & 7.20 \\
WavLM+MFA~\cite{guo2024audio} & 0.42 & 5.08 & 2.56 & - \\
Wav2Vec+VIB~\cite{eom2022anti} & 0.40 & 4.92 & - & - \\
OCKD~\cite{lu2024one} & 0.39 & 0.90 & 2.27 & 7.68 \\
GFL-FAD~\cite{wang2024genuine} & 0.25 & - & - & - \\
Wav2Vec+Linear~\cite{wang2023spoofed} & 0.22 & 3.63 & 3.65 & 16.17 \\
OC+ACS~\cite{kim2024one} & 0.17 & 1.30 & 2.19 & - \\
Wav2Vec+AASIST~\cite{tak2022automatic} & - & \underline{0.82} & 2.85 & - \\
Wav2Vec+AASIST2~\cite{zhang2024improving} & 0.15 & 1.61 & 2.77 & - \\
Wav2vec+Conformer+TCM~\cite{truong2024temporal} & - & 1.03 & 2.06 & - \\
Wav2vec+STJ-GAT+BLDL$^{\star}$~\cite{wu2024spoofing} & \textbf{0.06} & \textbf{0.56} & \underline{1.89} & - \\
\rowcolor{gray!10}LSR & 0.19 & 2.35 & 3.01 & 6.58 \\
\rowcolor{gray!10}LSR+LSA & 0.15 & 1.19 & 2.43 & \underline{5.92} \\ 
\rowcolor{gray!25}LSR+LSA$^{\star}$ & \underline{0.12} & 1.05 & \textbf{1.86} & \textbf{5.54} \\
\bottomrule
\end{tabular}
\begin{tablenotes}
\item $^{\star}$ with extra data augmentation.
\end{tablenotes}
\end{table}

To evaluate the overall performance of the proposed methods, we tested the system on four datasets and compared the results with those from the literature that used the same training dataset, as shown in Table~\ref{tab:compare}.
Across all datasets, LSR+LSA consistently outperforms LSR alone and often ranks among the top performers, highlighting the effectiveness of integrating latent space refinement with latent space augmentation. 
To further enhance the results, we applied additional data augmentation, which led to EERs of 0.12\% on \texttt{19LA}, 1.05\% on \texttt{21LA}, 1.86\% on \texttt{21DF}, and 5.54\% on \texttt{ITW}. This places our method on par with, or ahead of, the current state-of-the-art methods. Notably, our method focuses on refining and augmenting the latent space, which contrasts with recent approaches that focus on modifying the model architecture~\cite{truong2024temporal, wu2024spoofing}. These two strategies—latent space manipulation and architectural improvements—target different aspects of the problem and could potentially be combined for even better results. This highlights the flexibility and advantage of our method, as it enhances generalization without needing to alter the underlying model architecture.
In summary, the proposed LSR+LSA method consistently delivers strong results, matching or outperforming state-of-the-art performance across various datasets, demonstrating its robustness and effectiveness in generalizing across diverse deepfake detection tasks.

\begin{figure*}[ht]
    \centering
    \includegraphics[width=1\linewidth]{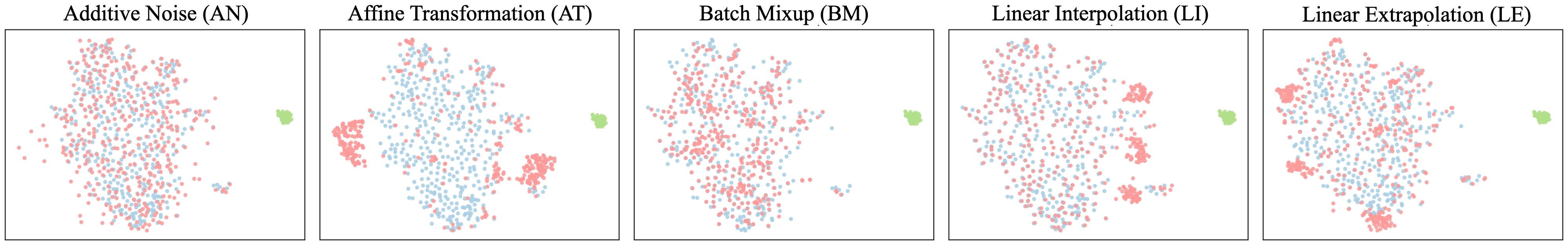}
    \caption{t-SNE visualization of the training dataset featuring various latent space augmentations. The green, blue, and red points represent the 2D projections of embeddings for the bonafide, spoof, and augmented spoof classes, respectively.}
    \label{fig:tsne}
\end{figure*}

\subsection{Ablation Study on Latent Space Refinement}
\vspace{-3mm}
\begin{table}[ht]
\centering
\caption{EER (\%) across datasets for systems trained with different loss configurations. Best results are in \textbf{bold}, and second-best results are \underline{underlined}.}
\label{tab:loss}
\begin{tabular}{l|cccc|c}
\toprule
Loss Configuration & \texttt{19LA} & \texttt{21LA} & \texttt{21DF} & \texttt{ITW} & Avg. \\ \midrule
WCE   & 0.30 & 2.64 & 4.74 & 8.09 & 3.94 \\
OC Softmax & 0.31 & \underline{1.60} & 4.06 & 7.86 & 3.46 \\
\rowcolor{gray!10}LSR & \underline{0.23} & \textbf{1.55} & \underline{3.22} & \underline{7.45} & \underline{3.11} \\
\rowcolor{gray!10}\quad w/o $\mathcal{L}_{inter}$ & 0.23 & 1.84 & 3.30 & 7.84 & 3.30 \\
\rowcolor{gray!10}\quad w/o $\mathcal{L}_{intra}$ & 0.27 & 2.62 & 4.02 & 7.75 & 3.67 \\
\rowcolor{gray!10}\quad w/o $\mathcal{L}_{intra}$, $\mathcal{L}_{inter}$ & 0.32 & 2.86 & 4.11 & 8.05 & 3.84 \\ 
WCE+LSR & \textbf{0.19} & 2.35 & \textbf{3.01} & \textbf{6.58} & \textbf{3.03} \\ \bottomrule
\end{tabular}
\end{table}

Table~\ref{tab:loss} presents the performance of various loss configurations during training. The baseline configuration uses weighted cross entropy (WCE) loss for binary classification, with OC Softmax~\cite{zhang2021one} included for comparison. Incorporating Latent Space Refinement (LSR) improves performance over both WCE and OC Softmax. 
We further examine the effects of LSR’s loss terms. Removing inter-class regularization results in minimal degradation, indicating that the core prototype-based loss sufficiently handles prototype separation. However, removing intra-class regularization significantly reduces performance, as this term is crucial for maintaining prototype diversity within the spoof class and preventing collapse. When both regularizations are removed, performance drops to baseline levels.
Additionally, combining LSR with WCE yields the best overall results. While WCE provides a solid foundation for binary classification, LSR refines the latent space to better capture variations in spoofed data. This combination leads to improved generalization across the datasets.

\begin{figure}
    \centering
    \includegraphics[width=0.48\textwidth]{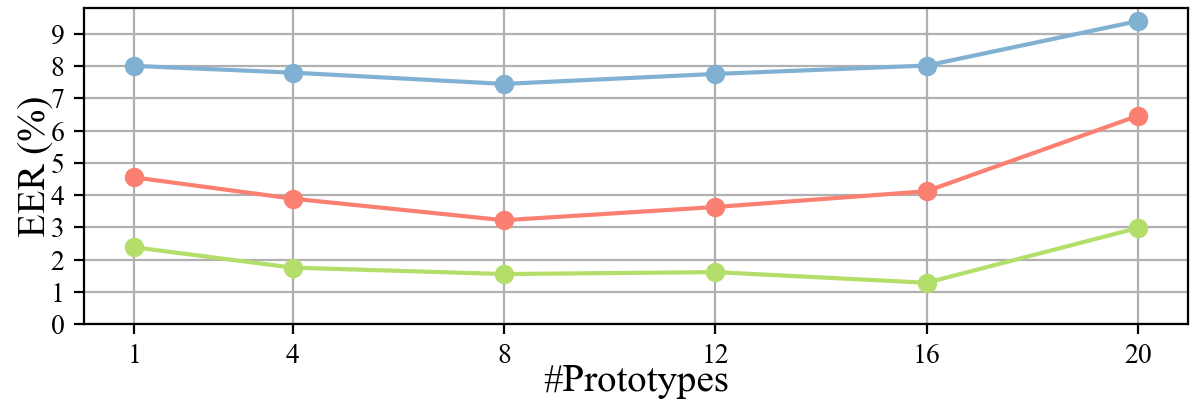}
    \caption{The effect of the number of spoofed prototypes on EER (\%) across different datasets (\texttt{21LA}, \texttt{21DF}, and \texttt{ITW}).}
    \label{fig:proto}
\end{figure}

Meanwhile, we evaluated the impact of the number of prototypes on performance, as shown in Fig.~\ref{fig:proto}. Increasing the prototypes from 1 to 8 improves performance, but further increasing to 16 shows diminishing returns. At 20 prototypes, performance declines, suggesting that too many prototypes can hinder generalization.

\subsection{Ablation Study on Latent Space Augmentation}
\vspace{-3mm}
\begin{table}[ht]
\centering
\caption{EER(\%) across datasets for systems trained with different latent space augmentation. Best results are in \textbf{bold}, and second-best results are \underline{underlined}.}
\label{tab:aug}
\begin{tabular}{l|cccc|c}
\toprule
Method & \texttt{19LA} & \texttt{21LA} & \texttt{21DF} & \texttt{ITW} & Avg.  \\\midrule
LSR & 0.19 & 2.35 & 3.01 & 6.58 & 3.03\\ 
\rowcolor{gray!10}\quad+LSA(AN) & \underline{0.16} & 1.67 & 2.85 & 6.17 & 2.71  \\
\rowcolor{gray!10}\quad+LSA(AT) & 0.19 & 1.62 & 2.57 & 6.69 & 2.77 \\ 
\rowcolor{gray!10}\quad+LSA(BM) & 0.21 & 1.65 & 2.86 & 6.61 & 2.93 \\
\rowcolor{gray!10}\quad+LSA(LI) & 0.23 & 1.92 & 2.65 & 7.05 & 2.96 \\ 
\rowcolor{gray!10}\quad+LSA(LE) & 0.18 & \underline{1.52} & \underline{2.54} & \underline{6.15} & \underline{2.60} \\ 
\rowcolor{gray!25}\quad+LSA(All) & \textbf{0.15} & \textbf{1.19} & \textbf{2.43} & \textbf{5.92} & \textbf{2.42} \\ \bottomrule
\end{tabular}
\end{table}

To assess the impact of different latent space augmentation methods, we conducted experiments for each method, as summarized in Table~\ref{tab:aug}, and visualized their effects using t-SNE in Fig.~\ref{fig:tsne}. Notably, since LI and LE rely on LSR prototypes, all systems were trained with LSR+WCE loss. 
Among the first three augmentations that are independent of the prototypes, AN and AT produced more dispersed and varied distributions, leading to better performance. In contrast, BM’s distribution remained closer to the original due to its mixup nature, which limited its effectiveness. 
For the prototype-dependent augmentations, LI, while beneficial, underperformed compared to the others, likely due to the consistent generation of challenging examples. LE, however, achieved the best results, as it effectively expanded the distribution into new regions of the latent space, offering a more balanced diversity.
Ultimately, combining all augmentation methods led to the most diverse latent space, resulting in the highest overall performance.

\begin{table}[ht]
\centering
\caption{Comparison of augmentation effects in input vs. latent space across datasets (EER \%).}
\label{tab:space}
\begin{tabular}{c|c|cccc|c}
\toprule
Method & Space & \texttt{19LA} & \texttt{21LA} & \texttt{21DF} & \texttt{ITW} & Avg.  \\ \midrule
None & - & 0.30 & 2.64 & 4.74 & 8.09 & 3.94\\ 
\rowcolor{gray!10}AN & input & 0.25 & 2.22 & 3.17 & 6.35 & 3.00  \\
\rowcolor{gray!10}AN & latent & 0.23 & 2.05 & 2.84 & 6.21 & 2.83 \\ 
AT & input & 0.27 & 2.43 & 3.44 & 6.81 & 3.24 \\
AT & latent & 0.25 & 2.03 & 2.91 & 6.72 & 2.98 \\ 
\rowcolor{gray!10}BM & input & 0.19 & 2.24 & 3.01 & 6.33 & 2.94 \\ 
\rowcolor{gray!10}BM & latent & 0.19 & 2.21 & 2.95 & 6.56 & 2.98 \\ 
\bottomrule
\end{tabular}
\end{table}

While we have demonstrated the effectiveness of augmentation in latent space, we were curious whether applying the same augmentations in the input space could yield comparable or even better results. To explore this, we conducted comparison experiments between augmentations applied in the input space versus the latent space, focusing on three methods that do not depend on latent prototypes or embeddings: AN, AT, and BM. All experiments were conducted using WCE loss without LSR. 
As shown in Table~\ref{tab:space}, applying augmentation, whether in the input or latent space, improves the baseline to some extent. 
For AN and AT, augmentations performed in the latent space consistently yield better results than those in the input space. This suggests that latent space augmentations may more effectively capture the underlying data distributions that the model needs to learn.
Interestingly, BM yields better results when applied in the input space than in the latent space. This outcome may be attributed to the nature of Mixup augmentation, which has been widely proven effective in various audio-related tasks when performed on the input data. The input space BM likely benefits from preserving more of the original data characteristics while still introducing beneficial variability.

\vspace{-2mm}
\section{Conclusions}
This paper presents a novel approach to enhance the generalization of audio deepfake detection systems by integrating Latent Space Refinement (LSR) and Latent Space Augmentation (LSA). LSR introduces multiple learnable prototypes to better capture the complex intra-class variability of spoofed audio, while LSA generates diverse representations in the latent space, further strengthening the model’s robustness. Extensive experiments on multiple datasets, including ASVspoof 2019 LA, ASVspoof 2021 LA, ASVspoof 2021 DF, and In-The-Wild, demonstrate that each of the proposed LSR and LSA can improve system significantly.

\section*{Acknowledgment}
This work was partially supported by the National Natural Science Foundation of China (NSFC) under Grants 62122050 and 62071288, and the Shanghai Municipal Science and Technology Commission under Grant 2021SHZDZX0102. Additional support was provided by the Pioneer R\&D Program of Zhejiang Province (No. 2024C01024) and the Ant Group Research Intern Program.

\bibliographystyle{IEEEbib}

\begin{thebibliography}{10}

\bibitem{yamagishi2021asvspoof}
Junichi Yamagishi, Xin Wang, Massimiliano Todisco, Md~Sahidullah, Jose Patino, Andreas Nautsch, Xuechen Liu, Kong~Aik Lee, Tomi Kinnunen, Nicholas Evans, et~al.,
\newblock ``Asvspoof 2021: accelerating progress in spoofed and deepfake speech detection,''
\newblock in {\em ASVspoof 2021 Workshop-Automatic Speaker Verification and Spoofing Coutermeasures Challenge}, 2021.

\bibitem{muller2022does}
Nicolas~M M{\"u}ller, Pavel Czempin, Franziska Dieckmann, Adam Froghyar, and Konstantin B{\"o}ttinger,
\newblock ``Does audio deepfake detection generalize?,''
\newblock {\em Interspeech}, 2022.

\bibitem{babu2021xlsr}
Arun Babu, Changhan Wang, Andros Tjandra, Kushal Lakhotia, Qiantong Xu, Naman Goyal, Kritika Singh, Patrick von Platen, Yatharth Saraf, Juan Pino, Alexei Baevski, Alexis Conneau, and Michael Auli,
\newblock ``Xls-r: Self-supervised cross-lingual speech representation learning at scale,''
\newblock {\em arXiv}, vol. abs/2111.09296, 2021.

\bibitem{radford2023robust}
Alec Radford, Jong~Wook Kim, Tao Xu, Greg Brockman, Christine McLeavey, and Ilya Sutskever,
\newblock ``Robust speech recognition via large-scale weak supervision,''
\newblock in {\em International conference on machine learning}. PMLR, 2023, pp. 28492--28518.

\bibitem{chen2022wavlm}
Sanyuan Chen, Chengyi Wang, Zhengyang Chen, Yu~Wu, Shujie Liu, Zhuo Chen, Jinyu Li, Naoyuki Kanda, Takuya Yoshioka, Xiong Xiao, et~al.,
\newblock ``Wavlm: Large-scale self-supervised pre-training for full stack speech processing,''
\newblock {\em IEEE Journal of Selected Topics in Signal Processing}, vol. 16, no. 6, pp. 1505--1518, 2022.

\bibitem{tak2022automatic}
Hemlata Tak, Massimiliano Todisco, Xin Wang, Jee-weon Jung, Junichi Yamagishi, and Nicholas Evans,
\newblock ``Automatic speaker verification spoofing and deepfake detection using wav2vec 2.0 and data augmentation,''
\newblock in {\em The Speaker and Language Recognition Workshop}, 2022.

\bibitem{zhang2021one}
You Zhang, Fei Jiang, and Zhiyao Duan,
\newblock ``One-class learning towards synthetic voice spoofing detection,''
\newblock {\em IEEE Signal Processing Letters}, vol. 28, pp. 937--941, 2021.

\bibitem{kim2024one}
Hyun~Myung Kim, Kangwook Jang, and Hoirin Kim,
\newblock ``One-class learning with adaptive centroid shift for audio deepfake detection,''
\newblock in {\em Interspeech 2024}, 2024, pp. 4853--4857.

\bibitem{park2019specaugment}
Daniel~S Park, William Chan, Yu~Zhang, Chung-Cheng Chiu, Barret Zoph, Ekin~D Cubuk, and Quoc~V Le,
\newblock ``Specaugment: A simple data augmentation method for automatic speech recognition,''
\newblock {\em arXiv preprint arXiv:1904.08779}, 2019.

\bibitem{tak2022rawboost}
Hemlata Tak, Madhu Kamble, Jose Patino, Massimiliano Todisco, and Nicholas Evans,
\newblock ``Rawboost: A raw data boosting and augmentation method applied to automatic speaker verification anti-spoofing,''
\newblock in {\em ICASSP 2022-2022 IEEE International Conference on Acoustics, Speech and Signal Processing (ICASSP)}. IEEE, 2022, pp. 6382--6386.

\bibitem{zhang2024cpaug}
Linjuan Zhang, Kong~Aik Lee, Lin Zhang, Longbiao Wang, and Baoning Niu,
\newblock ``Cpaug: Refining copy-paste augmentation for speech anti-spoofing,''
\newblock in {\em ICASSP 2024-2024 IEEE International Conference on Acoustics, Speech and Signal Processing (ICASSP)}. IEEE, 2024, pp. 10996--11000.

\bibitem{astrid2024targeted}
Marcella ASTRID, Enjie GHORBEL, and Djamila AOUADA,
\newblock ``Targeted augmented data for audio deepfake detection,''
\newblock in {\em 32nd European Signal Processing Conference (EUSIPCO 2024)}, 2024.

\bibitem{wang2023spoofed}
Xin Wang and Junichi Yamagishi,
\newblock ``Spoofed training data for speech spoofing countermeasure can be efficiently created using neural vocoders,''
\newblock in {\em ICASSP 2023-2023 IEEE International Conference on Acoustics, Speech and Signal Processing (ICASSP)}. IEEE, 2023, pp. 1--5.

\bibitem{wang2024can}
Xin Wang and Junichi Yamagishi,
\newblock ``Can large-scale vocoded spoofed data improve speech spoofing countermeasure with a self-supervised front end?,''
\newblock in {\em ICASSP 2024-2024 IEEE International Conference on Acoustics, Speech and Signal Processing (ICASSP)}. IEEE, 2024, pp. 10311--10315.

\bibitem{liu2018data}
Xiaofeng Liu, Yang Zou, Lingsheng Kong, Zhihui Diao, Junliang Yan, Jun Wang, Site Li, Ping Jia, and Jane You,
\newblock ``Data augmentation via latent space interpolation for image classification,''
\newblock in {\em 2018 24th International Conference on Pattern Recognition (ICPR)}. IEEE, 2018, pp. 728--733.

\bibitem{yan2024transcending}
Zhiyuan Yan, Yuhao Luo, Siwei Lyu, Qingshan Liu, and Baoyuan Wu,
\newblock ``Transcending forgery specificity with latent space augmentation for generalizable deepfake detection,''
\newblock in {\em Proceedings of the IEEE/CVF Conference on Computer Vision and Pattern Recognition}, 2024, pp. 8984--8994.

\bibitem{wang2020asvspoof}
Xin Wang, Junichi Yamagishi, Massimiliano Todisco, H{\'e}ctor Delgado, Andreas Nautsch, Nicholas Evans, Md~Sahidullah, Ville Vestman, Tomi Kinnunen, Kong~Aik Lee, et~al.,
\newblock ``Asvspoof 2019: A large-scale public database of synthesized, converted and replayed speech,''
\newblock {\em Computer Speech \& Language}, vol. 64, pp. 101114, 2020.

\bibitem{deng2019arcface}
Jiankang Deng, Jia Guo, Niannan Xue, and Stefanos Zafeiriou,
\newblock ``Arcface: Additive angular margin loss for deep face recognition,''
\newblock in {\em Proceedings of the IEEE/CVF conference on computer vision and pattern recognition}, 2019, pp. 4690--4699.

\bibitem{zhang2017mixup}
Hongyi Zhang, Moustapha Cisse, Yann~N Dauphin, and David Lopez-Paz,
\newblock ``mixup: Beyond empirical risk minimization,''
\newblock {\em arXiv preprint arXiv:1710.09412}, 2017.

\bibitem{jung2022aasist}
Jee-weon Jung, Hee-Soo Heo, Hemlata Tak, Hye-jin Shim, Joon~Son Chung, Bong-Jin Lee, Ha-Jin Yu, and Nicholas Evans,
\newblock ``Aasist: Audio anti-spoofing using integrated spectro-temporal graph attention networks,''
\newblock in {\em ICASSP 2022-2022 IEEE international conference on acoustics, speech and signal processing (ICASSP)}. IEEE, 2022, pp. 6367--6371.

\bibitem{pan2024attentive}
Zihan Pan, Tianchi Liu, Hardik~B. Sailor, and Qiongqiong Wang,
\newblock ``Attentive merging of hidden embeddings from pre-trained speech model for anti-spoofing detection,''
\newblock in {\em Interspeech 2024}, 2024, pp. 2090--2094.

\bibitem{oneata2023towards}
Octavian Pascu, Adriana Stan, Dan Oneata, Elisabeta Oneata, and Horia Cucu,
\newblock ``Towards generalisable and calibrated audio deepfake detection with self-supervised representations,''
\newblock in {\em Interspeech 2024}, 2024, pp. 4828--4832.

\bibitem{guo2024audio}
Yinlin Guo, Haofan Huang, Xi~Chen, He~Zhao, and Yuehai Wang,
\newblock ``Audio deepfake detection with self-supervised wavlm and multi-fusion attentive classifier,''
\newblock in {\em ICASSP 2024-2024 IEEE International Conference on Acoustics, Speech and Signal Processing (ICASSP)}. IEEE, 2024, pp. 12702--12706.

\bibitem{eom2022anti}
Youngsik Eom, Yeonghyeon Lee, Ji~Sub Um, and Hoi~Rin Kim,
\newblock ``Anti-spoofing using transfer learning with variational information bottleneck,''
\newblock in {\em Interspeech 2022}, 2022, pp. 3568--3572.

\bibitem{lu2024one}
Jingze Lu, Yuxiang Zhang, Wenchao Wang, Zengqiang Shang, and Pengyuan Zhang,
\newblock ``One-class knowledge distillation for spoofing speech detection,''
\newblock in {\em ICASSP 2024-2024 IEEE International Conference on Acoustics, Speech and Signal Processing (ICASSP)}. IEEE, 2024, pp. 11251--11255.

\bibitem{wang2024genuine}
Xiaopeng Wang, Ruibo Fu, Zhengqi Wen, Zhiyong Wang, Yuankun Xie, Yukun Liu, Jianhua Tao, Xuefei Liu, Yongwei Li, Xin Qi, Yi~Lu, and Shuchen Shi,
\newblock ``Genuine-focused learning using mask autoencoder for generalized fake audio detection,''
\newblock in {\em Interspeech 2024}, 2024, pp. 4848--4852.

\bibitem{zhang2024improving}
Yuxiang Zhang, Jingze Lu, Zengqiang Shang, Wenchao Wang, and Pengyuan Zhang,
\newblock ``Improving short utterance anti-spoofing with aasist2,''
\newblock in {\em ICASSP 2024-2024 IEEE International Conference on Acoustics, Speech and Signal Processing (ICASSP)}. IEEE, 2024, pp. 11636--11640.

\bibitem{truong2024temporal}
Duc-Tuan Truong, Ruijie Tao, Tuan Nguyen, Hieu-Thi Luong, Kong~Aik Lee, and Eng~Siong Chng,
\newblock ``Temporal-channel modeling in multi-head self-attention for synthetic speech detection,''
\newblock in {\em Interspeech 2024}, 2024, pp. 537--541.

\bibitem{wu2024spoofing}
Haochen Wu, Wu~Guo, Zhentao Zhang, Wenting Zhao, Shengyu Peng, and Jie Zhang,
\newblock ``Spoofing speech detection by modeling local spectro-temporal and long-term dependency,''
\newblock in {\em Interspeech 2024}, 2024, pp. 507--511.

\end{thebibliography}

\end{document}